# GENETIC CODE: FOUR-CODON AND NON-FOUR-CODON DEGENERACY

**Miloje M. Rakočević**

*Faculty of Science and Mathematics, Department of Chemistry, Višegradska 33, 18000 Niš, Serbia*
(E-mail: mirkovmiloje@gmail.com; www.rakocevcode.rs)

**Abstract**

In this work it is shown that 20 canonical amino acids (AAs) within genetic code appear to be a whole system with strict distinction in Genetic Code Table (GCT) into some different quantums: 20, 23, 61 amino acid molecules. These molecules distinction is followed by specific balanced atom number and/or nucleon number distinctions within those molecules. In this second version two appendices are added.

One of the most important question that has been made since creation of Genetic Code Table (GCT) (Crick, 1966, 1968) is the question of degeneration of the genetic code itself. Even then Rumer has noticed (Rumer, 1966) that the key aspect of degeneration could be found in the codon classification (and corresponding amino acids, AAs) according to the degeneration of type IV, when all four codons have the same aminoacidity meaning,[1] and to the degeneration of type non-IV or type I-II-III. Bearing in mind those facts, Scherbak has shown (Shcherbak, 1994) that classification of AAs into four-codon and non-four-codon AAs is followed by a specific balance of nucleons number, expressed by multiplying the number 037, which is named as Prime quantum (PQ).

It has been shown that this trend exists when calculation is performed with 8 AAs from eight quarter "cells" with codons of the same aminoacidity meaning and 15 AAs from eight also quarter "cells" but with codons of different aminoacidity meaning; all this, in total, is 23 amino acid molecules (Table 1 in relation to first area of Survey A1). On the other hand moguće je posmatrati skup od 61 AAs, korespondentno sa 61 amino acid kodonom (Table 2 in relation to second area of Survey A1). [Remark 1: It is important to notice the existence (here) two principles as significant and valid (Swanson, 1984; Rakočević, 2018b): the principle of continuity and principle of minimum change, which follow from the existance of specific proportions (30:30 = 1: 1; 30 : 60 = **1** : **2**; 9:16:25 = **3** : **4** : **5**).]

[1] For 4-letter alphabet the key distinction must be the degeneration of the type IV and non-IV; for the 5-letter alphabet – the type V and non-V, etc. Note in second version: With the knowledge that the genetic code is characterized by binary relationships [Pyrimidine as a single-ring and Purine as a two-ring molecule; the bonding in the U-A pair as weaker, and in the C-G pair as stronger hydrogen bonding; one and the second in relation to the binary relations in the set of 20 protein amino acids, as we have shown in previous works (Rakočević, 1998; Fig. 1, p. 284; 2018a; 2018b; 2019)], then we also can see that only four-letter alphabet, from the aspect of symmetry, can be correspondent to binarity (Rakočević, 2019, Table B5, p. 36).

| G | A | S | P | V | T | L | R | |
|---|---|---|---|---|---|---|---|---|
| 01 | 15 | 31 | 41 | 43 | 45 | 57 | 100 | 09 x 037 |
| 74 | 74 | 74 | 74 | 74 | 74 | 74 | 74 | **16 x 037** |
| 75 | 89 | 105 | 115 | 117 | 119 | 131 | 174 | 25 x 037 |

| C | I | S | L | N | D | Q | K | E | H | F | R | Y | M | W | |
|---|---|---|---|---|---|---|---|---|---|---|---|---|---|---|---|
| 47 | 57 | 31 | 57 | 58 | 59 | 72 | 72 | 73 | 81 | 91 | 100 | 107 | 75 | 130 | 30 x 037 |
| 74 | 74 | 74 | 74 | 74 | 74 | 74 | 74 | 74 | 74 | 74 | 74 | 74 | 74 | 74 | **30 x 037** |
| 121 | 131 | 105 | 131 | 132 | 133 | 146 | 146 | 147 | 155 | 165 | 174 | 181 | 149 | 204 | 60 x 037 |

**Table 1** The four-codon amino acids (above) and non-four-codon amino acids (down) as in Figure 1 in Shcherbak, 1994: nucleon number balances.

However, this paper deals with the fact that the determination law apllying the multiplying number 037 can be used even for the maximum possible aminoacidic distinction, when calculation is performed using all 61 amino acid molecules. Also, we show that Scherbak's system of multiplying number 037 (Table 1 in Shcherbak, 1994) is within one wider system of multiplying numbers 666 and 777 (Table 3), where it could be found that PQ 037 is generated from fraction 111/3. Bearing in mind those facts it is obvious that system with multiplying numbers 66 & 77 (with PQ 11/3) and 6 & 7 (with PQ 1/3) go before the system with multiplying numbers 666 and 777. Thus, those systems show that they are deteminants of classification into four-codon and non-four-codon AAs, if it is taken into account not only the number of nucleons (result 2 x 666 in Table 2, in relation to Table 4)[2] , but also the number of atoms (Tables 5 & 6).

[2] Table 4 (in a connection with Table 3) shows that, if in the coding system there are 8 x 8 codons, then must be 8 + 8 four-codon families with 8+15 = 23 amino acid molecules (neither 24 nor 25) [**8** + (8 – 1) = **15**].

| 4G | 4A | 4S | 4P | 4V | 4T | 4L | 4R | |
|---|---|---|---|---|---|---|---|---|
| 04 | 60 | 124 | 164 | 172 | 180 | 228 | 400 | 2 x 666<br>**36 X 037** |
| 296 | 296 | 296 | 296 | 296 | 296 | 296 | 296 | 64 x 037 |
| 300 | 356 | 420 | 460 | 468 | 476 | 524 | 696 | 100 x 037 |

| 2C | 3I | 2S | 2L | 2N | 2D | 2Q | 2K | 2E | 2H | 2F | 2R | 2Y | 1M | 1W | |
|---|---|---|---|---|---|---|---|---|---|---|---|---|---|---|---|
| 94 | 171 | 62 | 114 | 116 | 118 | 144 | 144 | 146 | 162 | 182 | 200 | 214 | 75 | 130 | **56 x 037** |
| 148 | 222 | 148 | 148 | 148 | 148 | 148 | 148 | 148 | 148 | 148 | 148 | 148 | 74 | 74 | 58 x 037 |
| 242 | 393 | 210 | 262 | 264 | 266 | 292 | 292 | 294 | 310 | 330 | 148 | 362 | 149 | 204 | 114 x 037 |

**Table 2.** The four-codon amino acids (above) and non-four-codon amino acids (down) as in Figure 1 in Shcherbak, 1994, all amino acids multiplied with coding codon number: nucleon number balances.

It can be seen, by comparing the Tables, still other different interesting balances, proportions and symetries. Thus, it can be found in Table 1 that determination using PQ 037 is realized in accordance with „the symmetry in the simplest case“ (Marcus, 1989). The symetry is for amino acid „heads“ deducing on proportion 1:2 (8 AAs molecules : 16 PQ and 15 AAs molecules : 30 PQ)[3], where the total quantum number is 46 (16+30). On the other hand, in Table 2 it can be found that in „bodies“, within AAs side chains there are 46±10 quantums (36 x 037 in singlemeaning and 56 x 037 in multimeaning cells).

The sense of determination by quantums „64“ and „58“, when dealing with nucleon number in the "heads" of AAs, is evident bearing in mind the following equation: **64** + $x$ = 2 x **61**, where $x$ = 58, and „64“ & “61” stand for known quantums in distinction of codon number, because 64 – 61 = 3, where quantum "3" regards to the number of stop codons. The total number of all quantums in Table 2 (100 + 114 = 214) corresponds on one side with the number of atoms in side chains of 20 AAs (2**1**4:2**0**4), while on the other side with the sum of the first four perfect numbers (2**1**4 : 2**3**4)(234 – 214 = 2**2**4). The change in the second position of number for 0, 1, 2, 3 shows that principes of continuity and minimum change stand here, as well[4]. [**Remark 2.** Notice the correspondence of quantums 214, 224 and 234 with the same quantums of atom number within 4 + 4 + 4 + 4 dinucleotides in Table 1 in Rumer, 1966; within 4 + 4 upper dinucleotides there are 116+108 = 224, and within 4 + 4 lower: 106 + 118 = 224 atoms. On the other hand, we have: 118 + 116 = 234 and 108 + 106 = 214 atoms. (As an interesting result is that, that the balance of atom number within dinucleotides is followed by a balance of atom number within 11 upper amino acid molecules (119) and 12 lower (120), all in Rumer's Table 1).]

[3] Result 16 PQ, i.e. 16 x 037 = 592 represents a half of the number 1184 which number is the first memeber of the second pair of friendly numbers. [Friendly number pairs: (220, 284), (1184, 1210), (17296, 18416) etc.]

| | | | |
|---|---|---|---|
| 1 | **037** | 666 | 777 |
| 2 | 074 | **1332** | 1554 |
| 3 | 111 | 1998 | **2331** |
| 4 | 148 | 2664 | 3108 |
| . . | . . . | . . . | . . . |
| 13 | 481 | **8658** | **10101** |
| 14 | 518 | 9324 | 10878 |
| . . | . . . | . . . | . . . |
| 16 | 592 | 10656 | 12432 |
| 17 | 629 | 11322 | 13209 |
| . . | . . . | . . . | . . . |
| 27 | **999** | 17982 | 20979 |

**Table 3.** The multiples of numbers 037, 666 and 777

[4] The sum of the first four perfect numbers is: 6+28+496+8128 = 8658 = 7770+0888 = 234 x 037; the 13th case of multiples of 666 in Table 3. (About the determination of genetic code with perfect and friendly numbers see in Rakočević, 1997, p. 60, or in: www.rakocevcode.rs). [Note in second version of this paper: Each of the last two columns in Table 3 (666 and 777) can be distribute into three new columns (6, 66, 666 and 7, 77, 777), as we have shown, for the first case, in {arXiv: 0903.4110 [q-bio.BM], Table A.1, p.21}, and as we have shown here for both cases, plus case 888, in Appendix B. In all this, it can be observed that in Table 3 there are particularly significant positions 02 and 13, which are at the beginning of the significant column within PSN (Figure A1, third column), showing the relationships of odd and even numbers on a decimal scale (Tables A1 to A6).]

| 8 + 7 = 15 | | 8 + 8 = 16 | | 8 + 9 = 17 |
|---|---|---|---|---|
| 8+ 15 = 23 | | 8+ 16 = 24 | | 8+ 17 = 25 |
| | | | | |
| 16 x 37 = 592 | | 16 x 37 = 592 | | 16 x 37 = 592 |
| 30 x 37 = 1110 | | 32 x 37 = 1184 | | 34 x 37 = 1258 |
| | | | | |
| 46 x 37 = 1702 | | 48 x 37 = 1776 | | 50 x 37 = 1850 |
| **56** x 37 = 2072 | | 58 x 37 = 2146 | | 60 x 37 = 2220 |
| **36** x 37 = **1332** | | 38 x 37 = 1406 | | 40 x 37 = 1480 |

**Table 4.** The quantums 8, 15 and 23 (above left) correspond with the patterns of 8 four-codon, 15 non-four-codon (and 23 in total) amino acids. The double values only in first column correspond to quantums in last column of Table 2 (36 as 46-10 and 56 as 46+10).

It is evident through the comparison of tabulated values in Tables 5 and 1, that in Table 5 determination regards the real 20 canonic AAs, while in Table 1, it regards with 23 AAs. To put it in other way, 6-codon AAs (L, S, R) play twice only in the case of nucleon number determination, but do not play in atom number determination. By comparing Table 5 and Table 6 it is shown that in the first case determination regards on the real 20 canonic AAs, while in the second case it regards on 61 AAs; in the first case with the determination through the multiplying number 6, while in the second case through the both multiplying numbers 6 and 66.

| G | A | S | P | V | T | L | R | | | | | |
| --- | --- | --- | --- | --- | --- | --- | --- | --- | --- | --- | --- | --- |
| 01 | 04 | 05 | 08 | 10 | 08 | 13 | 17 | | | | | 11 x 6 |
| 09 | 09 | 09 | 09 | 09 | 09 | 09 | 09 | | | | | **12 x 6** |
| 10 | 13 | 14 | 17 | 19 | 17 | 22 | 26 | | | | | 23 x 6 |
| C | I | N | D | Q | K | E | H | F | Y | M | W | |
| 05 | 13 | 08 | 07 | 11 | 15 | 10 | 11 | 14 | 15 | 11 | 18 | 23 X 6 |
| 09 | 09 | 09 | 09 | 09 | 09 | 09 | 09 | 09 | 09 | 09 | 09 | **18 x 6** |
| 14 | 22 | 17 | 16 | 20 | 24 | 19 | 20 | 23 | 24 | 20 | 27 | 41 x 6 |

**Table 5.** The four-codon amino acids plus six-codon amino acids (above) and non-four-codon amino acids (down): atom number balances.

| **4G** | **4A** | **4S** | **4P** | **4V** | **4T** | **4L** | **4R** | | | | | | | | |
| --- | --- | --- | --- | --- | --- | --- | --- | --- | --- | --- | --- | --- | --- | --- | --- |
| 04 | 16 | 20 | 32 | 40 | 32 | 52 | 68 | | | | | | | | **4 x 66**<br>44 x 6 |
| 40 | 52 | 56 | 68 | 76 | 68 | 88 | 104 | | | | | | | | 92 x 6<br>184 x 3 |
| **2C** | **3I** | **2S** | **2L** | **2N** | **2D** | **2Q** | **2K** | **2E** | **2H** | **2F** | **2R** | **2Y** | **1M** | **1W** | |
| 10 | 39 | 10 | 26 | 16 | 14 | 22 | 30 | 20 | 22 | 28 | 34 | 30 | 11 | 18 | **5 x 66**<br>55 x 6 |
| 28 | 66 | 28 | 44 | 34 | 32 | 40 | 48 | 38 | 40 | 46 | 52 | 48 | 20 | 27 | 197 x 3 |

**Table 6.** The four-codon amino acids plus six-codon amino acids (above) and non-four-codon amino acids (down), all amino acids multiplied with coding codon number: atom number balances.

In Conclusion we can say that all those tabulated results, reveals that all mentioned arithmetic regularities are of systematic character and stands for the whole code, as it is in the form of "standard genetic code". By this, those regularities comprise our hypothesis of complete genetic code (Rakočević, 2004), by which the standard genetic code would be systemic complete from the very beginning, as condition for life existence, while all irregularities from standard code are only the expression of the freedom rate in its minimum. But, there is still enigma, which Shcherbak pointed out in his paper, cited here, that is "the physical nature" of these arithmetical regularities "is so far not clear" (Shcherbak, 1993, p. 401). However, this should not be the trouble to reveal new arithmetical as well as algebraic regularities (Rakočević, 2018a, 2018b; 2019); also p-adic regularities (Dragovich and Dragovich, 2006), because it is the only way to investigate their not only physical and chemical but also systemic biological sense.

# Appendices

## **Appendix A.** The quantities of amino acid particles within PSN

The recently discovered fact that, like the chiral bonding (through mirror symmetry) between amino acid molecules, there is a chiral bonding between the characteristic numbers in the Periodic System of Numbers (PSN) and the number of atoms in the amino acid side chains (Rakočević, 2019, Fig. 2, p. 17 in relation to Table 3, p. 18), it made sense to investigate other possible correspondences of amino acid particles number and the numbers in PSN. Here are the first results obtained by following this trace, keeping in mind the splitting into four-codon and non-four-codon AAs.

**Survey A1.** Number of particles within four-codon and non-four-codon AAs

| **Table 1** | 8 Four-codon AAs | 15 Non-four-codon AAs | Remarks |
|---|---|---|---|
| Body of AAs | 09 x 037 = 333 | 30 x 037 = 1110 / 0111 | 333 + 1110 = 1443)[a] |
| Head of AAs | 16 x 037 = 592 | 30 x 037 = 1110/ 0111 | |
| **Total nucleons** | 25 x 037 = 925 | 60 x 037 = 2220/ 0222)[b] | 925 + 2220 = **3145**)[c] |
| 1443 + (23 x 74 = 1702) = 3145)[c] (592 + 1110 = 1702) (2072 – 1702 = 10 x 37) | | | 1332 = 2 x 666<br>2368 = 2 x 1184)[d] |
| **Table 2** | 32 Four-codon AAs | 29 Non-four-codon AAs | 2072 = 8 x 259)[e] |
| Body of AAs | 36 x 037 = **1332** | 56 x 037 = **2072** | 2072 – 1073 = 999)[f] |
| Head of AAs | 64 x 037 = **2368** | 58 x 037 = 1073 x 2 | 8658:2 = **4218**)[g] + 0111)[b] |
| **Total nucleons** | 100 x 037 = 3700 | 114 x 037 = **3145**)[c] +1073 | |
| [14 x 037 = 518] [14 x 037 = 518 (496 + 22] [ 114 x 037 = **4218**][g] | | | 4218 x 2 = 8436 |
| **Table 5** | 8 Four-codon AAs | 15 Non-four-codon AAs | |
| Body of AAs | 11 x 6 = 66 | 23 x 6 = 138 | 8436 = 8658 - 0222)[b] |
| Head of AAs | 12 x 6 = 72 | 18 x 6 = 108 | 264 = 4 x 66 |
| **Total atoms** | 23 x 6 = 138 | 41 x 6 = 123 x 2 | 394 – 10 = **384**)[h] |
| | | | 552 + **394** = 946 |
| **Table 6** | 32 Four-codon AAs | 29 Non-four-codon AAs | 496 + 946 = 1442)[k] |
| Body of AAs | 44 x 6 = 264 | 110 x 3 = **330**)[m] | **1443**)[a] – 1442 = 1 |
| Head of AAs | 48 x 6 = 288 | 87 x 3 = 261 | |
| **Total atoms** | 92 x 6 = 552 | 197 x 3 = **394** | |

**Legend to Survey A1**: (a) Nucleon number within the set of of 8 four-codon and 15 non-four-codon AAs in Standard genetic code (GC); (b) The mirror image symmetry; (c) Total nucleon number within the set of 23 AAs in Standard GC, which number together with its mirror image stands in a balance relation with the sum of the first four perfect numbers [3145 + 5413 = 8658 – 100] (6 + 28 + 496 + 8128 = 8658 = 7770 + 0888) (cf. Tabs B2 and B3); (d) The number 1184 as the first member within the first friendly number pair (1184 vs 1210); (e) The first of three Shcherbak's permutations (multiples of number 037) in determination of nucleon number within the set of 8 four-codon AAs; (f) Number 999 as last multiple in Shcherbak's Table of multiples of number 037 (cf. last row in Table 3); (g) Total nucleon number within 29 non-four-codon AAs in balance relation with the sum of the first four perfect numbers: 8658 – (2 x 4218 = 8436) = 0222. [On the other hand, the miror image of this result (2220 / 0222) is the same as the number od nucleons in 15 non-four-codon AAs within the set of 23 AAs (cf. last row in Table 1)]; (h) The number 384 as total number of atoms within whole molecules of 20 protein AAs. (About number 384 as a member of a specific harmonic system see in (Rakočević, 2011b, Tab. A.2, p. 839; here: Survey A3); (k) The sum of two permutations of the third perfect number; (m) The number 330 as the key determinant of atom number within standard GCT (Rakočević, 2004, Tab. 3a, p. 224; 2018a, Surv. 2, p. 37 and Comment 9, p. 40).

**Survey A2.** Relationships within six permutations of third perfect number

| | |
|---|---|
| (469 \| 964)<br>(496 \| 694)<br>(649 \| 946) | 469 + 964 + 694 + 649 = 4 x 694<br>4 x 694 = 2776 (1776 + 1000)<br>1776 = 2 x 888 (cf. Tab. B3) |
| 1442)[k] | 1442 + 2776 = 4218)[g] (½ 8436)<br>8436 = 8658 - 0222) (2220 / 0222)[b] |

**Legend to Survey A2**: Relationships within six permutations of third perfect number; (b) The mirror image symmetry; (g) The number 4218 as total nucleon number within 29 non-four-codon AAs in balance relation with the sum of the first four perfect numbers: 6 + 28 + 496 + 8128 = 8658; 8658 – (2 x 4218 = 8436) = 0222.

**Survey A3.** The relations of two Plato's progressions with Pythagorean musical note scale (Rakočević, 2011b, Tab. A.2, p. 839)

| Harmonic mean (h) | Arithmetic mean (m) |
|---|---|
| $h = \frac{2ab}{a+b}$ | $m = \frac{a+b}{2}$ |
| $1, \frac{4}{3}, \frac{3}{2}, 2, \frac{8}{3}, 3, 4, \frac{16}{3}, 6, 8$<br>$1, \frac{3}{2}, 2, 3, \frac{9}{2}, 6, 9, \frac{27}{2}, 18, 27$<br>$1, \frac{4}{3} \mid \frac{3}{2}, 2, \frac{8}{3} \mid 3, 4 \mid \frac{9}{2}, \frac{16}{3} \mid 6, 8 \mid 9, \frac{27}{2}, 18, 27$ | |
| $1, \frac{9}{8}, \frac{81}{64}\ \frac{4}{3} \mid \frac{3}{2}, \frac{27}{16}, \frac{243}{128}, 2$<br>$(3^2 \ \& \ 3^4 / 2^3 \ \& \ 2^6) \mid (3^{2+1} \ \& \ 3^{4+1} / 2^{3+1} \ \& \ 2^{6+1})$ | |
| 384 432 486 512 576 648 729 768<br>48 54 26 64 72 81 39 (384) | |

**Legend to Survey A3**: From (Rakočević, 2011b, Appendix, second Paragraph, p. 838: "The (Survey) represents the Plato's viewing how from two progressions ... follows a Pythagorean musicale tone scale. The order of four blocks as follows: 1. the formulas for harmonic and arithmetical means; 2. two progressions extended with harmonic and arithmetical means between two members in all steps, then combined; 3. first two members of progression with the quotient 2, the 1-2 interval, extended with harmonic and arithmetical means and then with additional two members through the rule given in row below in order to become a whole musicale tone scale (the quotient of all two neighbor numbers is 9/8, corresponding to a whole tone, except in two cases where it is 256/243, corresponding to a half-tone; 4. the reducing of fractions in the previous interval 1-2 in a common denominator, the number 384, wich number we can also find as a total number of atoms within 20 amino acid molecules."

| | ... | | | | | | | | | | |
|---|---|---|---|---|---|---|---|---|---|---|---|
| (-2) | ... | | | | | | | | | ... | -22 |
| (-1) | -21 | -20 | -19 | -18 | -17 | -16 | -15 | -14 | -13 | -12 | -11 |
| (0) | -10 | -09 | -08 | -07 | -06 | -05 | -04 | -03 | -02 | -01 | 00 |
| (1) | 01 | 02 | 03 | 04 | 05 | 06 | 07 | 08 | 09 | 10 | 11 |
| (2) | 12 | 13 | 14 | 15 | **16** | **17** | **18** | 19 | 20 | 21 | 22 |
| (3) | 23 | 24 | 25 | **26** | 27 | 28 | 29 | 30 | 31 | 32 | 33 |
| (4) | 34 | 35 | 36 | 37 | 38 | 39 | 40 | 41 | 42 | 43 | 44 |
| (5) | 45 | 46 | 47 | 48 | 49 | 50 | **51** | 52 | 53 | 54 | 55 |
| (6) | 56 | 57 | 58 | 59 | 60 | 61 | 62 | 63 | 64 | 65 | 66 |
| (7) | 67 | 68 | 69 | 70 | 71 | 72 | 73 | 74 | 75 | 76 | 77 |
| (8) | 78 | 79 | 80 | 81 | 82 | 83 | 84 | 85 | 86 | 87 | 88 |
| (9) | 89 | 90 | 91 | 92 | 93 | 94 | 95 | 96 | 97 | 98 | 99 |
| (A) | A0 | A1 | A2 | A3 | A4 | A5 | A6 | A7 | A8 | A9 | AA |
| (B) | B1 | B2 | B3 | B4 | B5 | B6 | B7 | B8 | B9 | BA | BB |

**Figure A1.** Periodic system of numbers (PSN) in decimal number system. Horizontaly: rows ("periods") and vertically: columns ("groups"); in red area: the same numbers through mirror symmetry, and in blue area also the same numbers through module 9; first blue column contains the relationships among odd (13, 35, 57, 79) as well as even numbers (02, 24, 46, 68) and/or (24, 46, 68, 8A) (A = 10; 8A = 90); the diagonal 11, 22, ..., 91 as a specific pattern for specific AAs arrangement, and the pattern 00-11-22 / 22-11-00 for other specific arrangement, are given in (Rakočević, 2019, Fig. 1, p. 6 and Fig. 2, p. 17 & Table 3, p. 18, respectively).

**Survey A4.** Relationships among rows and columns within PSN (I)

| |
|---|
| (01+11 = 12), (02+10 = 12), (03+09 = 12), (04+08 = 12), (05+07 = 12), (06+06 = 12) → 72 |
| (12+22 = 34), (13+21 = 34), (14+20 = 34), (15+19 = 34), (16+18 = 34), (17+17 = **34**) → **204**)[n] |
| (23+33 = 56), (24+32 = 56), (25+31 = 56), (26+30 = 56), (27+29 = 56), (28+28 = 56) → 336 |
| (34+44 = 78), (35+43 = 78), (36+42 = 78), (37+41 = 78), (38+40 = 78), (39+39 = 78) → 468 |
| (45+55 = 100), (46+54 = 100), (47+53 = 100), (48+52 = 100), (49+51 = 100), (50+50 = 100) → 600 |
| (56+66 = 122), (57+65 = 122), (58+64 = 122), (59+63 = 122), (60+62 = 122), (61+61 = 122) → 732 |
| (67+77 = 144), (68+76 = 144), (69+75 = 144), (70+74 = 144), (71+73 = 144), (72+72 = 144) → 864 |
| (78+88 = 166), (79+87 = 166), (80+86 = 166), (81+85 = 166), (82+84 = 166), (83+83 = **166**) → **996** |
| (89+99 = 188), (90+98 = 188), (91+97 = 188), (92+96 = 188), (93+95 = 188), (94+94 = 188) → 1128 |
| (12 + 188 = **200**), (34 + 166 = **200**), (56 + 144 = **200**), (78 + 122 = **200**), (100 + 100 = **200**)[p] |
| (72 + 1128 = 1200), (**204** + 996 = 1200), (336 + 864 = 1200), (468 + 732 = 1200), (600 + 600 = 1200)[q] |
| [1128 – 72 = 992 + 64), ( **996 – 204 = 992 – 200** ], |
| (864 – 336 = 992 – 464), (732 – 468 = 992 – 728), (600 – 600 = 992 – 992) |

**Legend to Survey A4:** Relationships within PSN in Figure 1: (n) The number 204 as the number of atoms in 20 protein amino acids, within their side chains; (p) The number 200 appears as determinant; (r) The number 1200 also appears as determinant. In the last output, all the relations shown are determined by the quantity of the number 204, and itself is determined by the double value of the third perfect number (2 x 496 = 992).

**Survey A5.** Relationships among rows and columns within PSN (II)

| | | | | |
|---|---|---|---|---|
| 12 + 188 = 200 | 188 = 496 – (330 – 22) | | 56 + 144 = 200 | 144 = 496 – (330 + 22) |
| 72 + 1128 = 1200 | 1128 = 496 + (500 + 132) | | 336 + 864 = 1200 | 864 = 496 + (500 – 132 ) |
| | | | | |
| 34 + 166 = 200 | 166 = 496 – 330)[m] | | 78 + 122 = 200 | 122 = 496 – (330 + 44) |
| **204** + 996 = 1200 | 996 = 496 + 500 | | 468 + 732 = 1200 | 732 = 496 + (500 – 264) |
| | | | | |
| **996 – 204 = 992 – 200** / 500 (010 / 101)[b] | | | 100 + 100 = 200 | 100 = 496 – (330 + 66) |
| | | | 600 + 600 = 1200 | 600 = 496 + (500 – 396) |

**Legend to Survey A5:** Relationships within PSN in Figure 1: (m) as in last sentence of Legend to Survey 1 [The number 330 as the key determinant of atom number within standard GCT (Rakočević, 2004, Tab. 3a, p. 224; 2018a, Surv. 2, p. 37 and Comment 9, p. 40)(cf. Legend to Table A6)]. In the last output, all the relations shown are determined by the quantity of the number 330, and itself is determined by the single value of the third perfect number (1 x 496 = 496). Notice the mirror symmetry between the numbers 2 and 5 in their binary notations (last row on the left (b); cf. legend to Fig. 2 in Rakočević, 2019).

**Table A1.** Odd numbers relations on the decimal scale (I)

| | | | | |
|---|---|---|---|---|
| 13 x 1 = 13 | 13 x 11 = 1**4**3 | 143 [a3] | 13 x 111 = 1**44**3 | **1443**)[a] |
| 35 x 1 = 35 | 35 x 11 = 3**8**5 | 385 [a3] | 35 x 111 = 3**88**5 | 3885)[r] |
| 57 x 1 = 57 | 57 x 11 = 5**C**7 | 627[a2] | 57 x 111 = 5**CC**7 | 6327 |
| 79 x 1 = 79 | 79 x 11 = 7**G**9 | 869 | 79 x 111 = 7**GG**9 | 8769 |
| A = 10, B = 11, C = 12, D = 13, E = 14, F = 15, G = 16, H = 17, K = 18 | | | | |
| [(13 + 79 = 35 + 57 = 92) (13 x **7** = 91) (92 – 91 = **1**] | | | | |
| [(143 + 869 = 385 + 627 = 1012) (13 x **77** = 1001) (1012 – 1001 = **11**] | | | | |
| [(**1443** + 8769 = 3885 + 6327 = 10212) (13 x **777** = 010101)[s] (10212 – 10101 = **111**] | | | | |
| (6 + 28 + 496 + 8128 = 8658) | (13 x 666 = 8658 = 010101[s] – **1443**) | | | |

**Legend to Table A1:** Relationships among odd numbers (13, 35, 57, 79) found in the first blue column of PSN (Fig. A1) trough multiplication with the numbers x, xx, xxx (x = 1). [The meaning of multiplication by these numbers, see in: Rakočević, 2018a, Survey 2, p. 37.] (a) Nucleon number within the set of of 8 four-codon and 15 non-four-codon AAs in Standard genetic code (GC); (a2) The number 627 as the "first half" (before the arithmetic mean) of the sum of all nucleons in the set of 20 AAs, within their side chains; (a3) The sum 143 + 385 = 628 - 100 as the unit-quantized-reduced "second half" (628) of the sum of all nucleons in the set of 20 AAs, within their side chains. By this the result 628 is behind the arithmetic mean [627 + 628 = 1255 (1443 – 1255 = 188 = L 57 + S 31 + R 11)]. [The sequence 143-385-627-628 appears to be a speciffic Fibonacci sequence: The previous two numbers give the follower of the following number, reduced by 1 in the third notation position, and increased by 1 in the first notation position: 143 + 385 = 628 - 100 and 627 + 001 = 628. As a hint more, the mirror symmetry also occurs in this "transaction": 100/001.] (r) The results of multiplication in the second row as follows: 35 = 5 x 7; 385 = 5 x 77; 3885 = 5 x 777; (s) Through the numbers 010101 and 8658 and their relations it follows that, of all the number systems, only the decimal system has a direct correspondence with the binary system, such as found on the binary tree of GC (Rakočević, 1998), thus with the Farey tree, with the Fibonacci series of numbers (and the Golden mean), and also with the sum of the first four perfect numbers, as well as with a unique mirror symmetry on the 6-bit binary tree.

**Table A2.** Odd numbers relations on the decimal scale (II)

| | | | | |
|---|---|---|---|---|
| 13 x 1 = 13 | 13 x 11 = 143 | 143 | 13 x 111 = 1443 | 1443 |
| 35 x 1 = 35 | 35 x 11 = 385 | 385 | 35 x 111 = 3885 | 3885 |
| 57 x 1 = 57 | 57 x 11 = 5C7 | 627 | 57 x 111 = 5CC7 | 6327 |
| 79 x 1 = 79 | 79 x 11 = 7G9 | 869 | 79 x 111 = 7GG9 | 8769 |
| A = 10, B = 11, C = 12, D = 13, E = 14, F = 15, G = 16, H = 17, K = 18 | | | | |
| [(13 + 79 = 35 + 57 = 92) (13 x 6 = 78) (92 – 78 = 14] = 7 x 2 | | | | |
| [(143 + 869 = 385 + 627 = 1012) (13 x 66 = 858) (1012 – 858 = 154] = 77 x 2 | | | | |
| [(1443 + 8769 = 3885 + 6327 = 10212) (13 x 666 = 8658 (10212 – 8658 = 1554] = 777 x 2 | | | | |
| (6 + 28 + 496 + 8128 = 8658) (13 x 666 = 8658 = 10101 – 1443) [1443 = 1554 – 111] | | | | |

**Legend to Table A2:** The new situation in relation to Table A1 is the multiplying 13 x 06 etc. instead of 13 x 07 etc.

**Table A3.** Even numbers relations on the decimal scale, including zero (I)

| | | | | |
|---|---|---|---|---|
| 02 x 1 = 02 | 02 x 11 = 022 | 022 | 02 x 111 = 0222 | 0222 / 2220)[b] |
| 24 x 1 = 24 | 24 x 11 = 264 | 264 | 24 x 111 = 2664 | 2664 = 4 x 666 |
| 46 x 1 = 46 | 46 x 11 = 4A6 | 506 | 46 x 111 = 4AA6 | 5106 |
| 68 x 1 = 68 | 68 x 11 = 6E8 | 748 | 68 x 111 = 6EE8 | 7548 (8658 - 1110)[t] |
| A = 10, B = 11, C = 12, D = 13, E = 14, F = 15, G = 16, H = 17, K = 18 | | | | |
| [(02 + 68 = 24 + 46 = 70) (02 x 07 = 14) (70 – 14 = 8 x (6 + 1)] = 56 | | | | |
| [(022 + 748 = 264 + 506 = 770) (02 x 077 = 154) (770 – 154 = 8 x (66 + 11] = 5B6 | | | | |
| [(0222+ 7548 = 2664 + 5106 = 7770) (02 x 0777 = 1554) (7770 – 1554 = 8 x (666 + 111] =5BB6 | | | | |
| (56 + 616 + 6216 = 6888 | | | 6888 = 8658 - 1770 [1770 – 770 = 1000] | |

**Legend to Table A3:** Relationships among even numbers (02, 24, 46, 68) found in the first blue column of PSN (Fig. A1) trough multiplication with the numbers x, xx, xxx (x = 1). (t) The number 1110 as the number of ncleons in 15 side chains as well as 15 heads of non-four-codon AAs (Survey A1, first and second row).

**Table A4.** Even numbers relations on the decimal scale, including zero (II)

| 02 x 1 = 02 | 02 x 11 = 022 | 022 | 02 x 111 = 0222 | 0222 |
|---|---|---|---|---|
| 24 x 1 = 24 | 24 x 11 = 264 | 264 | 24 x 111 = 2664 | 2664 |
| 46 x 1 = 46 | 46 x 11 = 4A6 | 506 | 46 x 111 = 4AA6 | 5106 |
| 68 x 1 = 68 | 68 x 11 = 6E8 | 748 | 68 x 111 = 6EE8 | 7548 |
| A = 10, B = 11, C = 12, D = 13, E = 14, F = 15, G = 16, H = 17, K = 18 | | | | |
| [(02 + 68 = 24 + 46 = **70**) (02 x **6** = 12) (**70** – 12 = 58 = 58) | | | | |
| [(022 + 748 = 264 + 506 = **770**) (02 x **66** = 132) (**770** – 132 = 5D8 = 638 | | | | |
| [(0222+ 7548 = 2664 + 5106 = **7770**) (02 x **666** = 1332) (**7770** – 1332 = 5DD8 = 6438 | | | | |
| (58 + 638 = 496 + 200<br>222 – 200 = 022 [220 / 022] | 6438 = 8658 – 2220 [2220 / 0222] | | | |

**Legend to Table A4:** The new situation in relation to Table A3 is the multiplying 2 x 06 etc. instead of 2 x 07 etc.

**Table A5.** Even numbers relations on the decimal scale, without zero (I)

| 24 x 1 = 24 | 24 x 11 = 264 | **264** | 24 x 111 = 2664 | 2664 |
|---|---|---|---|---|
| 46 x 1 = 46 | 46 x 11 = 4A6 | 506 | 46 x 111 = 4AA6 | 5106 |
| 68 x 1 = 68 | 68 x 11 = 6E8 | 748 | 68 x 111 = 6EE8 | 7548 |
| 8A x 1 = 90 | 8A x 11 = 8KA | **990** | 8A x 111 = 8KKA | 9990 |
| A = 10, B = 11, C = 12, D = 13, E = 14, F = 15, G = 16, H = 17, K = 18 | | | | |
| [(24+**90** = 46 + 68 = 114) (24 x **7** = 168) (168 – 114 = 54 (6 x **09**) | | | | |
| [(**264**[u] + **990** = 506 + 748 = 1254) (24 x **77** = 1848) (1848 – 1254 = **594**[u]] (6 x **099**) | | | | |
| [(2664 + **9990** = 5106 +7548 =12654) (24 x **777** = 18648) (18648 –12654 = 5994] (6 x **099**) | | | | |
| (54 + 594 + 5994 = 6642) (6642 = 8658 – **2016**) [Sum: 1 to 63 → 2016] | | | | |

**Legend to Table A5:** (u) The number 264 (8 x 033) and 594 as number of atoms in 61 amino acid molecules (in their side chains) within standard GCT (Rakočević, 2004, Tab. 3a, p. 224; 2018a, Surv. 2, p. 37 and Comment 9, p. 40). [Cf. Table A6.]

**Table A6.** Even numbers relations on the decimal scale, without zero (II)

| 24 x 1 = 24 | | 24 x 11 = 264 | 264 | | 24 x 111 = 2664 | 2664 = 4 x 666 |
|---|---|---|---|---|---|---|
| 46 x 1 = 46 | | 46 x 11 = 4A6 | 506 | | 46 x 111 = 4AA6 | 5106 |
| 68 x 1 = 68 | | 68 x 11 = 6E8 | 748 | | 68 x 111 = 6EE8 | 7548 |
| 8A x 1 = 90 | | 8A x 11 = 8KA | 990 | | 8A x 111 = 8KKA | 9990 |
| A = 10, B = 11, C = 12, D = 13, E = 14, F = 15, G = 16, H = 17, K = 18 | | | | | | |
| [(24 + 90 = 46 + 68 = 114) (24 x 6 = 144) (144 – 114 = 10 x 3 | | | | | | |
| [(264[u] + 990 = 506 + 748 = 1254) (24 x 66 = 1584) (1584 – 1254 = 10 x 33 = 330[u] | | | | | | |
| [(2664+ 9990 = 5106 + 7548 = 12654) (24 x 666 = 15984) (15984 – 12654 = 10 x 333] | | | | | | |
| (30 + 330 + 3330 = 3690 = 30 x 123 [1369 = 37 x 37] | | | | | | |

**Legend to Table A6:** (u) The number 264 (8 x 033) and 330 (10 x 33) as the number of atoms in 61 amino acid molecules (in their side chains: 264 + 330 = 594) within standard GCT (Rakočević, 2004, Tab. 3a, p. 224; 2018a, Surv. 2, p. 37 and Comment 9, p. 40). [Cf. Table A5.] From the First column within Table A1 to Table A6 follows the conclusion that only by the decimal number system, a series of natural numbers can be represented in the form of a Boolean logical square, and by the quaternary system in the form of a Boolean logical segment (Table A7). [Cf. Shcherbak, 1994, Table 1 ("Three-digit numbers of decimal system , multiples of 37") and Table 3 ("Sixty four triplets of the universal genetic code in the form of a complete set of three-digit numbers of the quaternary system ").]

**Table A7.** Natural numbers series as a Boolean logical space in correspondence with PSN

| (0) [00] | (1) [01] | (2) [10] | (3) [11] |
|---|---|---|---|
| 0, 1, 2 | 3, 4, 5 | 6, 7, 8 | 9, 10, 11 |
| 9, 10, 11 | 12, 13, 14 | 15, 16, 17 | 18, 19, 20 |
| 0, 1, 2 | 3, 4, 5 | 6, 7, 8 | 9, 10, 11 |
| … | | | |
| A logical segment (0, 1) in quaternary number system | | | |
| 0, 1, 2 | 3, 10, 11 | | |
| 3, 10, 11 | 12, 13, 20 | | |
| 0, 1, 2 | 3, 10, 11 | | |
| … | | | |

**Appendix B.** The multiples of the numbers x, xx, xxx (x = 6, 7, 8)

| Multiples of 01, 6, 66, 666, 037 | | | | |
|---|---|---|---|---|
| 01 | 6 | 66 | 666 | **037** |
| 162 = 216 – (2 x 27) | | | | |
| 27 | **162** | 1782 | 162 | **999** |
| 26 | 156 | 1716 | 17316 | **962** |
| 25 | 150 | 1650 | 16650 | **925** |
| **…** | | | | |
| 13 | 78 | 858 | **8658** | **481** |
| 12 | 72 | 792 | 7992 | **444** |
| 11 | 66 | 726 | 7326 | **407** |
| **…** | | | | |
| 03 | 18 | 198 | 1998 | **111** |
| 02 | 12 | 132 | 1332 | **074** |
| 01 | 6 | 66 | 666 | **037** |
| The 216 as Plato's number (6^3 = 216) | | | | |

**Table B1.** "The multiples of the numbers are presented in the first row. The 13th case is the sum of the first four perfect numbers (6 + 28 + 496 + 8128 = 8658)." (Rakočević, 2015, p. 61).

| Multiples of 01, 7, 77, 777, 037 | | | | |
|---|---|---|---|---|
| 01 | 7 | 77 | 777 | **037** |
| 189 = 216 – (1 x 27) | | | | |
| 27 | **189** | 2079 | 20979 | **999** |
| 26 | 182 | 2002 | 20202 | **962** |
| 25 | 175 | 1925 | 19425 | **925** |
| ... | | | | |
| 13 | 91 | 1001 | **10101** | **481** |
| 12 | 84 | 924 | 9324 | **444** |
| 11 | 77 | 847 | 8547 | **407** |
| ... | | | | |
| 03 | 21 | 231 | 2331 | **111** |
| 02 | 14 | 154 | 1554 | **074** |
| 01 | 7 | 77 | 777 | **037** |
| The 216 as Plato's number (6^3 = 216) | | | | |

**Table B2.** "The multiples of the numbers presented in the first row. The 13th case corresponds to the line of maximal changes (the change in each following step) on the binary tree (Rakočević, 1998)." (Rakočević, 2015, p. 62).

| Multiples of 01, 8, 88, 888, 037 | | | | |
|---|---|---|---|---|
| 01 | 8 | 88 | 888 | **037** |
| 216 = 216 ± (0 x 27) | | | | |
| 27 | **216** | 2376 | 23976 | **999** |
| 26 | 208 | 2288 | 23088 | **962** |
| 25 | 200 | 2200 | 22200 | **925** |
| ... | | | | |
| 13 | 104 | 1144 | **11544** | **481** |
| 12 | 96 | 1056 | 10656 | **444** |
| 11 | 88 | 968 | 9768 | **407** |
| ... | | | | |
| 03 | 24 | 264 | 2664 | **111** |
| 02 | 16 | 176 | 1776 | **074** |
| 01 | 8 | 88 | 888 | **037** |
| (3^3 = 27) (6^3 = 216) | | | | |

**Table B3.** "The multiples of the numbers are presented in the first row. The Plato's number 216 (the cube of number 6) appears as the last result in column of number 8." (Rakočević, 2015, p. 63).